\documentclass[pdftex]{article}
\usepackage{icrctc07}

\title{A Deconvolution technique for VHE Gamma-ray Astronomy, and its application to the morphological study of shell-type supernova remnants}
\shorttitle{Deconvolution technique for VHE Gamma-Ray Astronomy}
\authors{G. Maurin$^{1}$, A. Djannati-Ata\"{i}$^{2}$, P. Espigat$^{2}$.}
\shortauthors{G. Maurin and et al}
\afiliations{$^1$DSM/DAPNIA - Direction des Sciences de la Matiere, Laboratoire de Recherche sur les lois fondamentales de l'Univers, CEA Saclay, 91191 Gif-Sur-Yvette Cedex, France\\ $^2$ Laboratoire APC - 10, rue Alice Domon et Léonie
Duquet 75205 Paris Cedex 13 - FRANCE}
\email{gilles.maurin@cea.fr}

\abstract{Deconvolution algorithms have been used successfully for optimization/restoration/deblurring of astronomical images in a variety of wavelengths, especially in the optical band (e.g., for HST). We present here an iterative \textsc{Richardson-Lucy} type method designed for treatment of images obtained with the H.E.S.S. array of ground-based gamma-ray telescopes. Its application to shell-type supernova remnant images yields refined details relevant for the study of correlations with other wavelengths, and hence for interpretation in terms either of hadronic or leptonic origin of the observed VHE gamma-ray emission.}

\begin{document}
\maketitle

\section{Introduction}
All images or maps obtained with a telescope are a statistical realization of the real source distribution degraded by the instrument response and with a certain noise: a point source appears thus with a spread distribution called the point spread function (PSF). So, in order to correcte these effects, we have to subtract noise and to deconvolve the map with the PSF. Several methods are used to do this treatment but they are not adapted for VHE gamma-ray astronomy in which the noise is comparable to source signal and in which we have less statistics. In this paper, we adapt the \textsc{Richardson-Lucy} method, based on a maximum likelihood (searching for the most probable source distribution) to these conditions. We then apply this method to study the shell of the sypernova remnant RXJ1713.7-3946.
\section{The classical \textsc{Richardson-Lucy Method}}
The \textsc{Richardson-Lucy} method algorithm was developped from Bayes's theorem [\cite{Richardson,Lucy}]. Its approach consists of constructing the conditionnal propability density relationship:
\begin{equation}
p(O|N_{s}) = \frac{p(N_{s}|O)p(O)}{p(N_{s})}
\label{equ1}
\end{equation}
where $p(N_{s})$ is the probability of the observed image $N_{s}(x,y)$ and $p(O)$ is the probability of the real image $O(x,y)$, \textit{i.e. the source distribution}, over all possible image realizations. The bayes solution is found by maximizing the right part of the equation, \textit{i.e. by searching the source distribution $O(x,y)$ which maximizes the probability (\ref{equ1})}. 

The term $P(N_{s})$ has a constant value and can be ignored during the maximization process, while we can assume a uniform probability density for the prior $p(O)$. So the maximum probability corresponds to the maximum likelihood over $O(x,y)$:
\begin{equation}
ML(O) = \max_{O}{p(N_{s}|O)}
\label{equ2}
\end{equation}

The probability $P(N_{s}|O)$ depends on the object and the noise. In our case, they follow Poisson distributions. The probability $p(N_{s}|O)$ is then given by:
\begin{equation}
p(N_{s}|O)=\prod_{x,y}\frac{\bar{N}_{s}(x,y)^{N_{s}(x,y)}e^{-\bar{N}_{s}(x,y)}}{N_{s}(x,y)!}
\label{equ3}
\end{equation}
where $\bar{N}_{s}$ is the convolution beetween the source distribution $O(x,y)$ and the normalized point-spread function $Psf(x,y)$:
\begin{equation}
\bar{N}_{s}(x,y)=O(x,y)*Psf(x,y)
\label{equ3b}
\end{equation}

The maximum likelihood can be computed by taking the derivative of the logarithm:
\begin{equation}
\frac{\partial\ln{p(N_{s}|O)}}{\partial O}=0
\label{equ4}
\end{equation}

Using Picard iteration [\cite{Picard}], we have:
\begin{eqnarray*}
O^{(n+1)}(x,y)=\left[ \frac{N_{s}(x,y)}{\bar{N}^{(n)}_{s}(x,y)}*Psf(x,y)\right]\times\\
O^{(n)}(x,y) =\alpha^{(n)}(x,y)O^{(n)}(x,y)
\end{eqnarray*}

\begin{equation}
\Rightarrow \alpha^{(n)}(x,y) = \frac{N_{s}(x,y)}{O(x,y)*Psf(x,y)}*Psf(x,y)
\end{equation}

The maximum likelihood is then obtained when all the coefficients $\alpha^{n}(x,y)$ converge to 1. This algorithm, commonly used in astronomy, intrinsically applies the positivity constraint of the solution and conserves flux. But this method doesn't work when the noise becomes important. Indeed, in this case, it creates after a few number of iterations high frequency structures which are not physical. We thus propose to modify this method in order to work with gamma-ray maps which present an important background.

\section{Deconvolution of maps with noise}
Commonly in gamma-ray astronomy, a background measurement is made for each observation. There are in fact several methods in order to do this study (ring background, template background...) but a noise map $N_{off}(x,y)$ is always obtained and can be sustracted to the $N_{on}(x,y)$ map to compute the excess map. In fact, the image data $\bar{N}_{on}$ is assumed to be due to the source distribution $O(x,y)$ convolued by the PSF and to a noise $\bar{N}_{off}(x,y)$ by:
\begin{eqnarray*}
\bar{N}_{on}(x,y)=&(O(x,y)*Psf(x,y))A(x,y)\\&+\bar{N}_{off}(x,y)
\end{eqnarray*}
where $A(x,y)$ is the gamma acceptance-ray in the feld of view of the telescopes. 

Then we can construct a conditionnal propability density relationship, in order to take into account the signal and noise distributions:
\begin{eqnarray*}
p(\bar{N}_{on}|N_{on})~p(\bar{N}_{off}|N_{off})=~~~~~~~~~~~~~~~~~~~~~~~~~~~~~~~~~~~\\
\frac{p(N_{on}|\bar{N}_{on})p(\bar{N}_{on})}{p(N_{on})}\times\frac{p(N_{off}|\bar{N}_{off})p(\bar{N}_{off})}{p(N_{off})}~~~~~
\end{eqnarray*}

Using the same assumptions than in the classical method, the solution is obtained for the maximum likelihood over $O(x,y)$ and over $\bar{N}_{off}$. These maximum likelihoods can be computed by taking the derivative of the logarithm:
\begin{eqnarray*}
\frac{\partial\ln{\left[p(N_{on}|\bar{N}_{on})~p(N_{off}|\bar{N_{off}})\right]}}{\partial O}=0
~~~~~\\~~~~\rm{and}~~~~
\frac{\partial\ln{\left[p(N_{on}|\bar{N}_{on})~p(N_{off}|\bar{N_{off}})\right]}}{\partial \bar{N}_{off}}=0
\end{eqnarray*}

The maximum likelihood over $\bar{N}_{off}$, is obtained for:
\begin{equation}
\bar{N}_{off}(x,y)=N_{off}(x,y)
\label{equ9}
\end{equation}
So $\bar{N}_{off}$ can be directly remplaced by $N_{off}(x,y)$ for the maximum likelihood over $O(x,y)$. We find then a new iteration coefficient $\alpha^{n}(x,y)$ in Picard iteration given by:
\begin{eqnarray*}
\alpha^{(n)}(x,y) =~~~~~~~~~~~~~~~~~~~~~~~~~~~~~~~~~~~~~~~~~~~~~~~~~~~~~~~~~~\\ \frac{\frac{N_{on}(x,y)A(x,y)}{(O^{(n)}(x,y)*Psf(x,y))A(x,y)+N_{off}(x,y)}* Psf(x,y)}{A(x,y)*Psf(x,y)}
\end{eqnarray*}

In this way, we take into account both noise and acceptance while allowing negative values for the source distribution. To avoid this problem, we can impose $O(x,y)$ positive in each pixel during the Picard steps. This condition brakes however the photometry, but we have show on simulations that this effect is very weak ($<1$\%). 

\section{When does the modified \textsc{Richarsond-Lucy} method converge
?}
There are several criteria used to test the convergence of the deconvolution method
(bootstrap method [\cite{Bootstrap}], comparison with realistic simulations...). Here
we propose a simple approch where we stop the deconvolution algorithm, pixel-by-pixel,
when the coefficient $\alpha^{n}(x,y)$ reaches 1 with a choosen limit. This limit
depends on the PSF and the bin size. So we have to adjust it to the map that we want
to analyse. The next paragraph presents this work on the supernova remnant:
RXJ1713.7-3946.

\section{Application to the supernova remnant: RXJ1713.7-3946}
The figure \ref{rxj1713} (left) presents the excess map of RXJ1713.7-3946 obtained by the H.E.S.S. telescopes [\cite{hess}]. The map binning is of $0.02^\circ$ and the PSF is modelized by simulations through a 2-dimensionnal Gaussian of $\sigma\approx0.1^\circ$. 

\subsection{Estimation of the optimal iteration number}
\begin{figure}
\begin{center}
\includegraphics[width=0.50\textwidth]{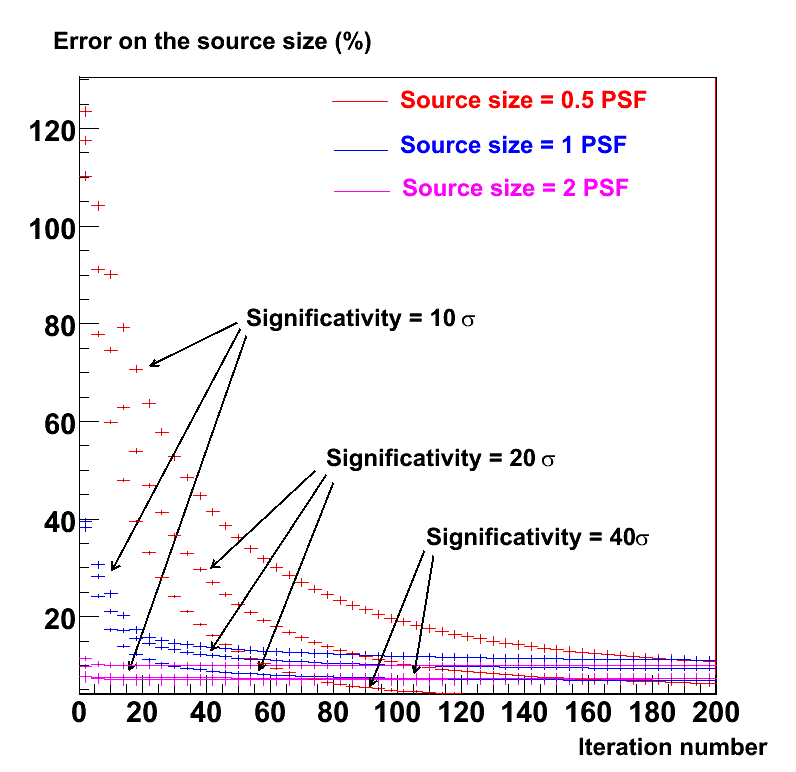}
\end{center}
\caption{Source size error as a function of intensity for Gaussian- and ring-shaped sources with different sizes (from 0.5 to 2\,PSF size).}
\label{Result}
\end{figure}

The supernova remnant shell projected on the sky appears as a fraction of a ring with
several hot spots. To estimate the optimal iteration number, we simulate Gaussian- and
ring-shaped sources with different sizes (from 0.5 to 2\,PSF size) and for several
intensity values (yielding a significance between 10 to 40$\sigma$) and with the same
binning as data. For these simulations, we seach for the best iteration number with a
limit on $\alpha^{n}(x,y)$ (see previous paragraph) and evaluate the spatial resolution by
fitting the source size after deconvolution. The plots given in \ref{Result} presents
these resolutions with a limit of 1\%.

This method makes converge the deconvolution to a limit of resolution
depending on the significance and the source size. The higher the significance or the bigger the source, the
better the resolution is. These tests show also that the deconvolution is a slow process after several
iterations. So, if we don't apply exactly the optimal iteration number, it doesn't modify significantly the deconvolued map.

\subsection{The deconvolution of RXJ1713.7-3946}
\begin{figure*}[th]
\begin{center}
\includegraphics[width=0.90\textwidth]{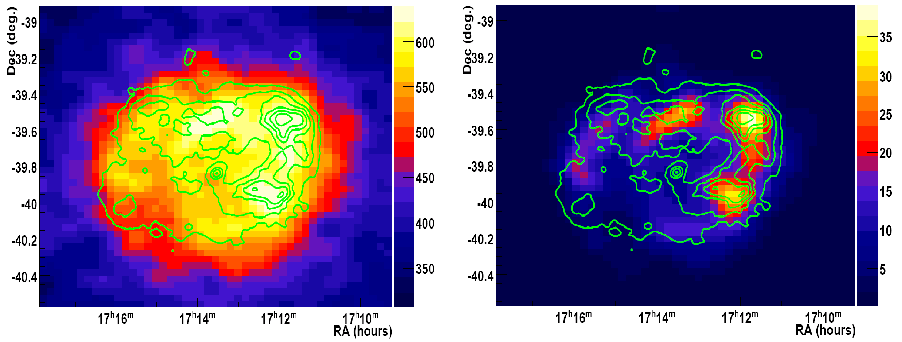}
\end{center}
\caption{Excess (left) and deconvolved (right) map of RXJ1713.7-3946 observed by H.E.S.S. The excess map is smoothed by a gaussian of $0.04^\circ$ in order to see the structure of the supernova shield. In both maps, the ASCA experiment (1-3 \textrm{keV}) contours [\cite{ASCA}] are overdrawn.}
\label{rxj1713}
\end{figure*}

The RXJ1713.7-3946 map through H.E.S.S. observations yields a significance of $30\sigma$. The source
morphology is close to a fraction of a ring of width $\approx0.2^\circ$. Applying the deconvolution with 120 iterations (according to fig.\ref{Result}) we obtain the map drown in fig.\ref{rxj1713} right.

\begin{figure}
\begin{center}
\includegraphics[width=0.48\textwidth]{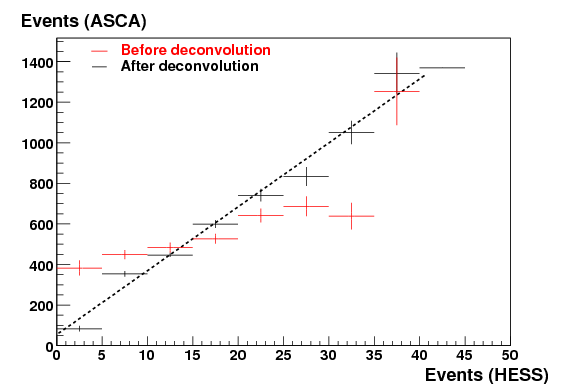}
\end{center}
\caption{Correlation between the \textrm{TeV} emission from the H.E.S.S. map and the X-rays (\textrm{1-3 keV}) from the ASCA map [\cite{ASCA}].}
\label{Correl}
\end{figure}

The deconvolution improves clearly the correlation between the \textrm{TeV} emission and the X-ray emission (1-3 \textrm{keV}) as measured by ASCA [\cite{ASCA}]. Figure \ref{Correl} presents the improvement by plotting the TeV emission as a function of X-rays emission from the ASCA map for each bin in the east part of the map. These correlations are important to constrain models.\\

In leptonic models, synchrotron emission of electrons generates X-ray and inverse Compton with photon target (optical star light, CMB, IR) generates the TeV emission. We then expect strong correlations. In the hadronic modeld, Tev emission is due to the $\pi^0$ decay generated by the interactions between accelerated hadrons and nuclei in the site. These interaction generate electron which could be responsible to the X-ray emission. But, at the moment, this process has to be check by modelization.   

\section{Conclusion and future prospects}
Our method, based on a classical algorithm for treatment of images in the optical band, allows us to improve the angular resolution of gamma-ray astronomy by making a statistical correction of the instrumental spread. This analysis can be already used to study the morphology of extended sources. In a futur step, the background treatment should be improved during the deconvolution process and the limits of the algorithm should be estimated.

\bibliography{icrc0615}

\begin{thebibliography}{1}

\bibitem{Bootstrap}
H.~Bi and G.~Boerner.
\newblock When does the richardson-lucy deconvolution converge ?
\newblock {\em astroph/9409093}, 1994.

\bibitem{hess}
The~H.E.S.S. Collaboration.
\newblock Primary particle acceleration above 100 tev in the shell-type
  supernova remnant rx j1713.7-3946 with deep h.e.s.s. observations.
\newblock volume 449, pages 223--242, 2006.

\bibitem{Picard}
E.~Isaacson and H.~Keller.
\newblock Analysis of numerical methods.
\newblock 1966.

\bibitem{ASCA}
K.~Koyama.
\newblock Discovery of non-thermal x-rays from the northwest shell of the new
  snr rx j1713.7-3946: The second sn 1006?
\newblock {\em PASJ}, 49:L7, 1997.

\bibitem{Lucy}
L.B. Lucy.
\newblock {\em An iterative technique for the rectification of observed
  distribution}, volume~79.
\newblock 1974.

\bibitem{Richardson}
W.H. Richardson.
\newblock {\em Bayesian-based iterative method of image restoration},
  volume~62.
\newblock 1972.

\end{thebibliography}
\bibliographystyle{plain}

\end{document}